\documentclass[prl,twoside,showpacs,twocolumn,floatfix]{revtex4}

\usepackage{amssymb}
\usepackage{amsfonts}
\usepackage{amsmath}
\usepackage[dvips]{graphicx}
\usepackage{color}
\newcommand{\dd}{\mathrm{d}}
\begin{document}

\title{Aging and effective temperatures near a critical point}
\author{S. Joubaud, B. Percier, A. Petrosyan and S. Ciliberto}
\affiliation{ Universit\'e de Lyon, Laboratoire de Physique, Ecole
Normale Sup\'erieure de  Lyon, CNRS , 46, All\'ee d'Italie, 69364 Lyon CEDEX 07, France}
\date{\today}

\begin{abstract}
The orientation fluctuations of the director of a liquid
crystal (LC) are measured after a quench near the Fr\'eedericksz
transition, which is a second order transition driven by an
electric field. We report experimental evidence that,  because of
the critical slowing down, the LC presents several properties of an
aging system after the quench, such as power law scaling in times of correlation and response functions. During this
slow relaxation, a well defined effective temperature, much larger
than the heat bath temperature,  can be measured  using the
fluctuation dissipation relation.
\end{abstract}


\maketitle The  characterization of the thermodynamic properties
of out of equilibrium and of slow relaxing systems is an important
problem of great current interest, which is studied both
theoretically and experimentally. In this framework an important
and general question, which has been analyzed only theoretically
is the relaxation of a system, which is rapidly quenched exactly
at the critical point of a second order phase
transition~\cite{Peter,Berthier,Abriet2004,Gambassi,Gambassi1}. {
Because of the well known divergence of the relaxation time and of
the correlation length the system presents a very rich dynamic,
which has been named ``aging at the critical point''. Indeed this
relaxation dynamic presents several properties, which are
reminiscent of those observed during the aging of spin
glasses~\cite{Peter,Berthier,Gambassi,Gambassi1,Abriet2004,Cugliandolo2002,Vincent}
: the mean quantity has an algebraic decay and  the correlation
and response functions exhibit a power law scaling in time.
Another important feature analyzed during the aging at a critical
point concerns the equilibrium relation between response and
correlation, {\em i.e.} the fluctuation dissipation theorem (FDT),
which is not necessarily satisfied in an out of equilibrium
system. This statement is relevant in the context of spin glass
aging where a well defined (in the thermodynamic sense)  effective
temperature $T_{\rm eff}$ can be obtained  using the so called
fluctuation dissipation relation (FDR). This relation leads to the
introduction of the fluctuation dissipation ratio, $X$, between
correlation and response
functions~\cite{Cugliandolo1997,Cugliandolo2002}. Deviations of
$X$ from unity ($X=1$ is the equilibrium value) may quantify the
distance from equilibrium~\cite{Cugliandolo1997}. Equivalently an
effective temperature $T_{\rm eff}=T/X$, larger than the heat bath
temperature $T$, can be defined. Those ideas have been transferred
to the aging at critical point, where for certain variables the
FDR can be interpreted as a well defined effective temperature
$T_{\rm eff}$~\cite{Peter, Berthier}.

The study of these analogies between the aging at the critical point
and the aging of spin glasses is important because it allows to
give new insight on the role of the quenched disorder on the above
mentioned features and on the common mechanisms producing them.
However these theoretical studies have been performed mainly on spin
models and  most importantly the reduced control parameter
$\epsilon$ has been set exactly equal to  zero. Thus one is
interested to know how general these predictions are  and also one
may wonder whether those predictions  can be observed  in an
experimental system where the exact condition $\epsilon=0$ can
never be reached.

The purpose of this letter is to analyze, within this theoretical
framework, an experiment on the relaxation dynamic close to the
critical point of a liquid crystal instability, which  is, at a
first approximation, described by a Ginzburg-Landau equation and
has the main features of a second order phase transition. The main
result of our investigation is that after a quench close to the
critical point  the effective temperature $T_{\rm{eff}}$, measured
from FDR, has a well defined value larger than the thermal bath
temperature.

This kind of experimental test is  important because the
properties of $X$ and $T_{\rm eff}$ have attracted much interest,
since they suggest that a generalized statistical mechanics can be
defined for a broad class of non equilibrium phenomena. Several
theoretical models or numerical simulations show such a behavior
\cite{Cugliandolo2002,Berthier2007}. However in experiments the
results are less clear and the relevance for real materials of the
$T_{\rm{eff}}$ as defined by FDR is still an open question, which
merits investigation~\cite{Israeloff,Bellon2001,Ocio,Abou03}. For
example, there are important differences, which are not
understood, between supercooled fluids \cite{Israeloff},
polymers\cite{Bellon2001}, gels\cite{Abou03} and spin
glasses\cite{Ocio}.} The point is that on this subject it is very
difficult to find simple theoretical models, which can be directly
compared with experiments. Thus the experimental study of the
relaxation dynamics close to the critical point is very useful in
this sense.


The system of our interest is the Fr\'eedericksz transition of a
liquid crystal (LC), subjected to an electric field
$\vec{E}$~\cite{DeGennes,Oswald}. In this system, we measure the
variable $\zeta$,  which is the spatially averaged alignment of
the LC molecules, whose local direction of alignment is defined by
the unit pseudo vector $\vec{n}$. Let us first recall the general
properties of the Fr\'eedericksz transition. The system under
consideration is a LC confined between two parallel glass plates
at a distance $L = 9$~$\mu$m. The inner surfaces of the confining
plates have transparent Indium-Tin-Oxyde (ITO) electrodes, used to
apply the electric field. Furthermore the plate surfaces, are
coated by a thin layer of polymer mechanically rubbed in one
direction. This surface treatment causes the alignment of the LC
molecules in a unique direction parallel to the surface (planar
alignment),{\em i.e.} all the molecules have the same director
parallel to the $x$-axis\cite{cell}. The cell is next filled by a
LC having a positive dielectric anisotropy
(p-pentyl-cyanobiphenyl, 5CB, produced by Merck). The LC is
subjected to an electric field perpendicular to the plates, by
applying a voltage $V$ between the ITO electrodes, {\em i.e.} ${E}
= V/L$. To avoid the electrical polarization of the LC, we apply
an AC voltage at a frequency $f_V = 1$~kHz ($V = \sqrt{2} V_0
\cos(2\pi f_V t)$)~\cite{DeGennes, Oswald}. More details on the
experimental set-up can be found in
ref.\cite{Joubaud_upon,joubaud2008}. When $V_0$ exceeds a
threshold value $V_c$ the planar state becomes unstable and  the
LC molecules, except those anchored to the glass surfaces, try to
align parallel to the field, {\em i.e.} the director, away from
the confining plates, acquires a component parallel to the applied
electric field ($z$-axis). This is the Fr\'eedericksz transition
whose properties are those of a second order phase
transition~\cite{DeGennes,Oswald}. For $V_0$ close to $V_c$ the
motion of the director is characterized by its angular
displacement $\theta$ in $xz$-plane ($\theta$ is the angle between
the $x$ axis and $\vec{n}$), whose space-time dependence has the
following form : $\theta = \theta_0(x,y,t) \sin\left(\frac{\pi
z}{L}\right)$~\cite{DeGennes, Oswald, SanMiguel1985}. If
$\theta_0$ remains small then its dynamics, neglecting the $(x,y)$
dependence of $\theta_0$, is described by a Ginzburg-Landau
equation :
\begin{eqnarray}
\tau_0 \frac{{\rm d} \theta_0}{{\rm d} t}=  \epsilon \ \theta_0 -
{1 \over 2} (\kappa+\epsilon+1) \theta_0^3 +\eta
\label{momentum_equation}
\end{eqnarray}
where $\epsilon = \frac{V_0^2}{V_c^2} - 1 $ is  the reduced
control parameter. The characteristic time $\tau_0$  and constant
$ k $  depend on the LC material properties.  $\eta$ is a thermal
noise delta-correlated in time \cite{SanMiguel1985}.

We define the variable $\zeta$ as the spatially averaged alignment
of the LC molecules and more precisely :
\begin{equation}
\zeta = \frac{2}{L}\frac{1}{\mathcal{A}}\int\hspace*{-0.2cm}
\int_{\mathcal{A}} \dd x\dd y \int_0^{L}(1-n_x^2)\dd z
\label{eq:def_zeta}
\end{equation}
where $\mathcal A=\pi D_0^2/4$ is the area,  in the ($x$, $y$)
plane,  of the measuring region of diameter $D_0$, which is about
$2$ mm in our case. If $\theta_0$ remains small, $\zeta$ takes a
simple form in terms of $\theta_0$, i.e. $n_x = \cos(\theta)$ and
$ \ \zeta = 1/\mathcal{A}\int\hspace*{-0.1cm}\int_{\mathcal{A}}
\dd x\dd y \theta_0^2$. It has been  shown that $\zeta$ is
characterized by a mean value $\langle \zeta \rangle \propto
\epsilon$, by a divergent relaxation time $\tau = \tau_o/\epsilon
$ and by fluctuations, which have a Lorentzian
spectrum~\cite{Galatola, joubaud2008}. The measure of the variable
$\zeta$ relies upon  the anisotropic properties of the LC, {\em
i.e.} the cell is a birefringent plate whose local optical axis is
parallel to $\vec{n}$. This optical anisotropy can be precisely
measured using a polarization interferometer~\cite{Bellon02},
which has a signal to noise ratio larger than 1000 (see ref.
\cite{Joubaud_upon,joubaud2008} for details).

\begin{figure}
\centering
\includegraphics[width=0.9\linewidth]{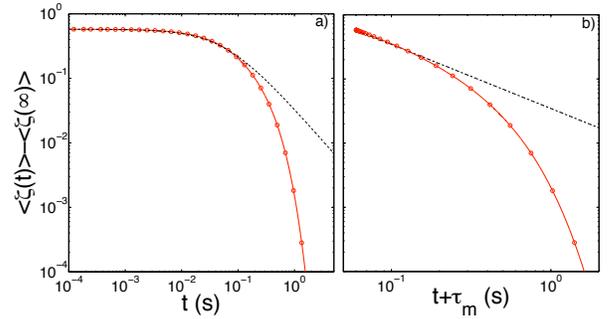}
\caption{a) Mean relaxation of the system, $\log \langle \zeta(t)
\rangle$, as a function of time after a quench ($\circ$) from
$\epsilon_1=0.3$ to $\epsilon_0=0.01$. The continuous line
represents the theoretical response based on
eq.~\ref{momentum_equation} with $\langle \zeta \rangle \simeq
\psi_0^2$. b) Mean relaxation
 of the system as a function of {  the reduced  time $(t+\tau_m)$}.
in both figures the dashed lines represents the theoretical
relaxation at $\epsilon=0$.} \label{repfit}
\end{figure}

In this letter, we consider the dynamics of $\zeta$ as a function
of time after a quench from $\epsilon=\epsilon_1>0.3$ to $\epsilon
= \epsilon_0 <<0.1$. The fact that the control parameter is the
electric field allows one to have extremely fast quenches,
typically $1$~ms. As the typical relaxation time $\tau$ of $\zeta$
is about $10s$, at $\epsilon_0\simeq0$, this means that we can
follow the out of equilibrium dynamics for about four orders of
magnitude in time, which is really comparable to what is done in
real glass experiments after a temperature quench,  where  the
dynamics is usually followed for about 3  or 4 decades in time.
The advantage here is that this relaxation dynamics, which lasts
only a few seconds, allows us to repeat the experiment several
times and to perform an ensemble average (indicated by $<\cdot>$)
of the measured quantities. We consider first the specific case of
a quench from $\epsilon_1=0.3$ to $\epsilon_0=0.01$. The typical
mean value $<\zeta(t)>$ of $\zeta$ after this quench is plotted in
fig.~\ref{repfit} as a function of time $t$ ($t=0$ is the time
when the quench has been performed). This mean dynamics of $\zeta$
is obtained by repeating the same quench $7000$ times. The
behavior of $<\zeta>$ remains constant for a certain time and then
slowly relaxes (see fig.~\ref{repfit}a). Above a characteristic
time, which is about $0.22$~s in our case, the relaxation becomes
exponential.

To understand this behavior, we decompose the dynamics of $\theta_0$ in a mean dynamics after
the quench and its fluctuations, i.e. $\theta_0=\psi_0(t) +\delta
\theta_0$, where $\psi_0(t) =\langle \theta_0(t) \rangle$. The
dynamics of $\Psi_0$ obtained from the analytical solution of
eq.\ref{momentum_equation} is shown in fig.\ref{repfit} as a
continuous line, which perfectly agrees with the measured
dynamics. The experimental data and the  solution of
eq.\ref{momentum_equation} have
two well distinguish limits  : for $t\gg \tau \equiv
\tau_0/(2\epsilon)$, the relaxation is exponential ($\tau\simeq
0.22s$ at $\epsilon_0=0.01$) ; for $t\ll
\tau\equiv\tau_0/(2\epsilon_0)$, the dynamics of $\psi_0$ is
almost algebraic : $\psi_0(t) = \left[ {(\kappa+1)\over \tau_0}
\left(t +\tau_m \right) \right]^{-1/2}$ with
$\tau_m=\frac{\tau_0}{(\kappa+1)\psi_0(0)^2}$. This behavior,
plotted
 in fig.~\ref{repfit} as a dashed line,  is  identical to the relaxation
at the critical point~\cite{Gambassi}.{  Thus the system should
present aging phenomena in the range $t<\tau$, which is the
interval where the dashed line follows the experimental data in
fig.~\ref{repfit}.}

{  In order to measure $T_{\rm eff}$ as defined by the FDR, we
need to measure both the correlation function $C_{\theta}(t,t_w)
\equiv \langle \ \delta \theta_0(t) \ \delta \theta_0(t_w)\rangle$
of the thermal fluctuations $\delta \theta_0$ of $\theta_0$ and
the response function $\chi(t,t_w)$ of $\theta_0$, at time $t$  to
a perturbation, given at time $t_w$, to its conjugated variable
$\Gamma$, with $t_w<t$. More precisely, from
ref.\cite{Cugliandolo1997,Cugliandolo2002} the effective
temperature is given by: }
\begin{equation}
\chi(t,t_w)= {X(t,t_w) \over k_B T }
[C_{\theta}(t,t)-C_{\theta}(t,t_w)] \label{eq:FDR}
\end{equation}
where $k_B$ is the Boltzmann constant,T the heat bath temperature,
i.e. $T_{\rm eff}=T/X(t,t_w)$.

To compute $\chi(t,t_w)$ and $C_{\theta}(t,t_w)$ in our experiment
one has to consider that the measured variable is $\zeta$.{  The
relationship between the fluctuations of $\zeta$ and $\theta_0$
has been discussed in details in ref.\cite{Joubaud_upon}, thus  we
recall here only the useful results}. As the area of the measuring
beam is much larger than the correlation length, the global
variable measured by the interferometer is $\zeta=
\frac{1}{\mathcal A}\iint \theta_0^2 \dd x \dd y \simeq \psi_0^2 +
2 \psi_0 \delta \theta_0$. Thus the mean value of $\zeta$ is
$\langle \zeta(t)\rangle = \psi_0^2(t)$ and the fluctuations of
$\delta \zeta$ of $\zeta$ can be related to the fluctuations of
$\theta_0$ : $\delta \zeta(t)  = 2\psi_0(t)\delta \theta_0(t)$.

\begin{figure}
\centering
\includegraphics[width=1\linewidth]{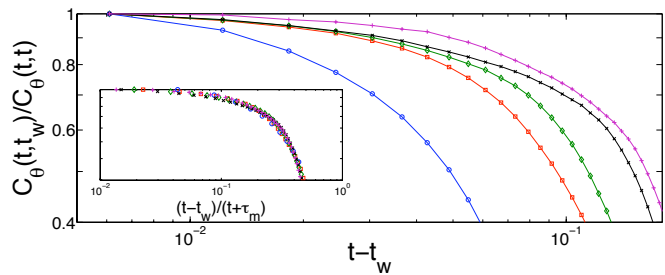}
\caption{a) Correlations functions $C_\theta(t,t_w)$ as a function
of $t-t_w$ for different fixed $t = 0.06 \ s$ ($\circ$), $0.2\ s$
($\Box$), $0.26 \ s$ ($\diamond$), $0.28 \ s$ ($\times$) and
$0.31\ s$ ($+$) after the same quench of fig.~\ref{repfit}. Inset:
The correlation functions have a simple master curve obtained by
plotting $C_\theta(t,t_w)/C_\theta(t,t_w)$ versus
$(t-t_w)/(t+\tau_m)$. Notice that $(t+\tau_m)$ is the reduced time
used in fig.\ref{repfit}} \label{fig:correl}
\end{figure}

The autocorrelation function of $\theta_0$ is obtained using the
values of $\psi_0(t)$
 and $\psi_0(t_w)$, {\em i.e.} $C_{\theta}(t,t_w) =
 \langle \zeta(t)\zeta(t_w)\rangle/(4\psi_0(t)\psi_0(t_w))$.
The $C_{\theta}(t,t_w)$, measured at various fixed $t$, are
plotted as function of $t-t_w$ in fig.\ref{fig:correl}.
We see that a simple scaling of the correlation function can be
obtained by plotting them as a function of $(t-t_w)/(t+\tau_m)$.
Thus in agreement with theoretical predictions, our system
presents, during its slow relaxation after the quench close to the
critical point, algebraic decay of mean quantities and  the
scaling of correlation functions, which are very  similar to the
main features of  spin glass aging (see for example
\cite{Cugliandolo2002,Vincent}).

\begin{figure}
\centering{
\includegraphics[width=0.85\linewidth]{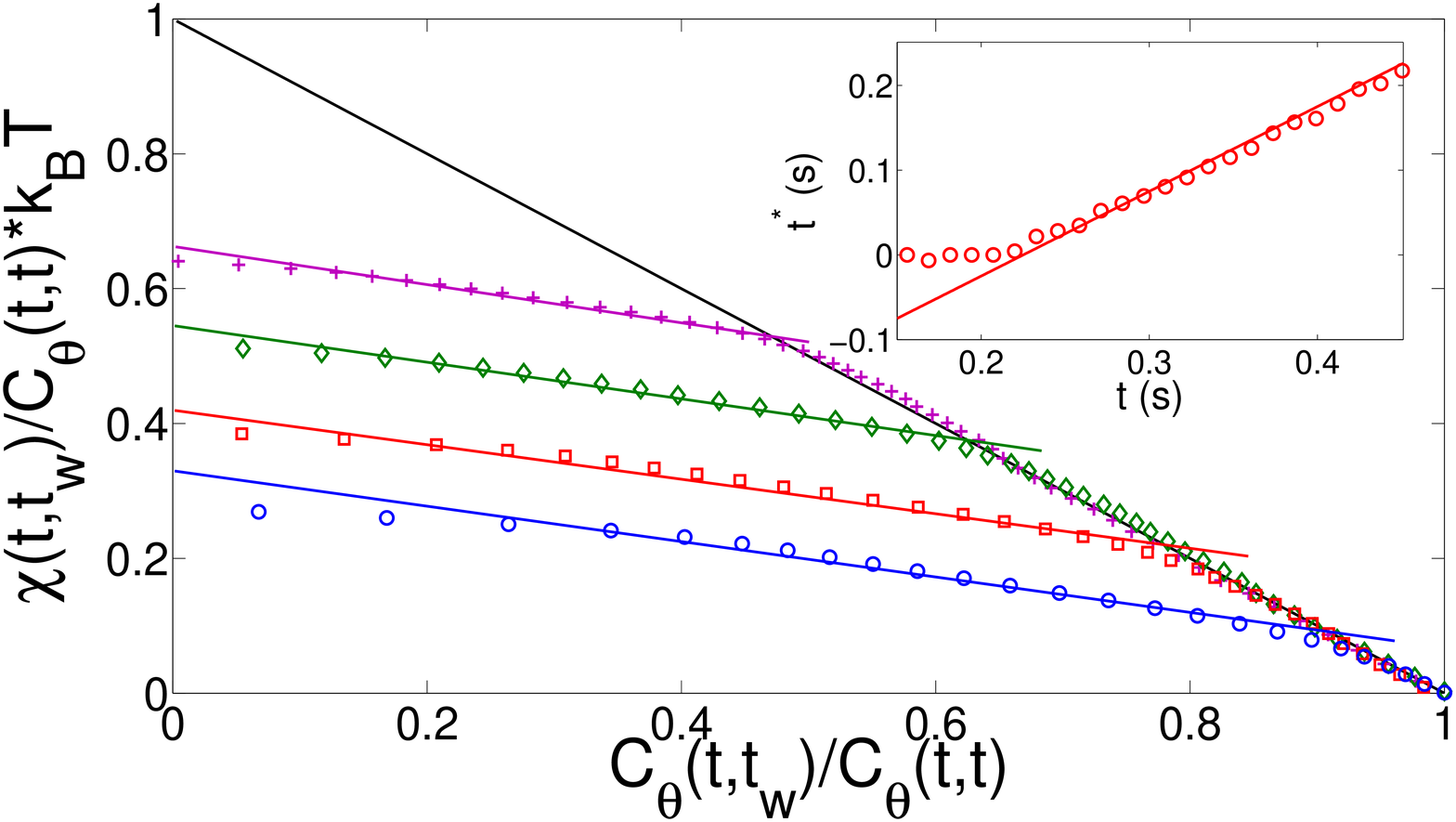}}
\caption{a) Parametric plot of integrated response versus
correlation for   $t = 0.4$~s ($\circ$), $0.5$~s ($\Box$), $0.6$~s
($\diamond$) and $0.7$~s ($+$). Continuous line are linear fits.
Dark line represents the FDT at equilibrium. Inset : the
characteristic time $t^*$ is plotted as a function of $t$.}
\label{FDTs_t_fixed}
\end{figure}

The response function is obtained  by applying a small change of
the voltage $V_0$,
 which modifies the control parameter, {\em i.e.} $\epsilon = \epsilon_0+\delta \epsilon$.
In ref.\cite{Joubaud_upon} we have shown that the external torque
$\Gamma_{ext}$, i.e. the conjugated variable of $\theta_0$,
associated to $\delta \epsilon $ is equal to :
\begin{equation}\Gamma_{\rm ext} = 2B \ \delta \epsilon \ \psi_0(t) \
\left(1-\frac{\psi_0(t)^2}{2}\right)
\end{equation}
We separate $\theta_0(t)$ into the average part $\psi_0(t)$,
solution of eq.~\ref{momentum_equation}, with $\eta=0$, and a
deviation $\Delta(t)$ due to $\delta \epsilon$ : $\theta_0(t)
=\psi_0(t)+\Delta(t)$. We define the linear response function
\mbox{$R(t,t_w)=<\Delta> /\Gamma_{\rm ext}$} of $\theta_0$ to
$\Gamma_{ext}$ using a Dirac delta function for $\delta \epsilon$
at an instant $t_w$, {\em i.e.} $\delta \epsilon(t) \propto
\delta(t-t_w)$. In the experiment the Dirac delta function is
approximated by a triangular function of amplitude $\delta
\epsilon_0$ and duration $\tau_r \simeq 2ms <<\tau$, specifically:
$\delta \epsilon=\delta \epsilon_0(1- 2 |t-t_w|/\tau_r$ for
$|t-t_w|<\tau_r/2$ and $\delta \epsilon=0$ for $|t-t_w|>\tau_r/2$,
with $\tau_r/\tau \simeq 10^{-4}$. The measured quantity is the
response  $R_{\zeta, \delta \epsilon}=<\Delta \zeta>/\delta
\epsilon$, of $\zeta(t)$ to $\delta \epsilon(t_w)$. As $\Delta
\zeta (t)= 2\psi_0(t) \Delta $ it follows that the linear response
of $\theta_0$ to $\Gamma_{ext}$ is:
\begin{equation}
R(t,t_w) = \frac{R_{\zeta, \delta \epsilon}(t,t_w)}{4B\
\psi_0(t_w) \ \psi_0(t)\ \left(1-\frac{\psi_0(t_w)^2}{2}\right)}
\label{eq:chi}
\end{equation}
where $B=A\pi^2K_1/(4L)$, and $K_1$ a LC elastic constant. Thus
inserting in eq.\ref{eq:chi} the measured values of $R_{\zeta,
\delta \epsilon}(t,t_w)$, of $\psi_0(t)$ and of $\psi_0(t_w)$ we
can measure $R(t,t_w)$ and by numerical integration we finally
obtain $\chi(t,t_w)= \int_{t}^{t_w} R(t,t') \dd t'$. The accuracy
of this procedure has been checked at equilibrium in
ref.\cite{Joubaud_upon}, where we have shown that the measured
$C_{\theta_0}$ and $\chi(t,t_w)$ verify FDT.

The integrated response is related to the correlation by the FDR
relations (see eq.~\ref{eq:FDR}). Due to the non-equilibrium
process, it is not equivalent whether the parameter is the waiting
time $t_w$ or the observation time $t$ (see
ref.\cite{Berthier2007} for a detailed discussion on this point).
Applying the correct procedure \cite{Cugliandolo1997,Berthier}, we
keep the time $t$ constant and we vary $t_w$, i.e. $0<t_w<t$, and
we repeat the procedure at various $t$. Following
ref.\cite{Cugliandolo1997}, we study the FDR during the relaxation
process by plotting the integrated response $\chi$ as a function
of $C_{\theta}$. These FDR plots can be seen in
fig.~\ref{FDTs_t_fixed} for four characteristic values of $t$. {
When the system is in equilibrium the FDR plot is a straight line
with slope $-1/(k_BT)$ (continuous line in
fig.~\ref{FDTs_t_fixed}). The curves in fig.~\ref{FDTs_t_fixed}
for each $t$ are composed by two straight lines and are remarkably
similar to those predicted in theoretical models
\cite{Cugliandolo1997} and seen in experiments \cite{Ocio} of spin
glasses aging.  When $t_w$ is close to $t$, i.e. large values of
$C(t,t_w)$, the FDT is satisfied, i.e. the experimental points are
on the continuous line in fig.~\ref{FDTs_t_fixed}. For $C(t,t_w)$
smaller than a value $C^{*}$, which depends on $t$, the FDT is not
satisfied. However $\chi(t,t_w)$ remains linear in $C(t,t_w)$ but
the slope, which does not depend on $t$,  is much smaller than its
value at equilibrium. This means that the slow modes of the
systems have a very well defined $T_{eff}$, independent of $t$,
higher than the temperature of the bath. Precisely we find, for
$C(t,t_w)<C^{*}$, $X= T/T_{\rm eff}=0.31$, which is remarkably
close to the asymptotic value $X=0.305$ analytically
\cite{Gambassi1} and numerically  \cite{Berthier} estimated for
the small wave vector modes of an Ising model quenched at the
critical point using a different quenching procedure. This analogy
is important because our measurement, being an integral over the
measuring volume, is more sensitive to the fluctuations of the
long wavelength modes. In fig.\ref{FDTs_t_fixed} we see that
$C^{*}$ is a decreasing function of $t$.  Let us define $t^{*}$ as
the characteristic time associated with $C^{*}$. In the inset of
fig.~\ref{FDTs_t_fixed}, we have plotted the value of $t^{*}$ as a
function of $t$. We find that   $t^*\simeq0$ for $t\leq
\tau=0.22s$. This indicates that, in agreement with
ref.\cite{Berthier,Gambassi,Gambassi1}, the violation occurs for
all times at $\epsilon_0=0$. In our case it is cut by the
exponential relaxation, which starts at $t\simeq\tau$. For
$t>\tau$, $t^*=t-\tau$ is linear in $t$, which also agrees with
the theoretical picture of the FDR \cite{Cugliandolo1997}. Indeed
$t^{*}$ separates the equilibrium part and the aging one and the
ratio $t^{*}/t=1-\tau/t$, for $t>\tau$, defines how large is the
equilibrium interval with respect to the total time. Notice that
at $\epsilon_0=0$, $\tau=\infty$ and  the equilibrium interval
does not exist \cite{Berthier}. Thus the values of $\epsilon_0$
and $\epsilon_1$ affect only the amplitude of the region where the
out of equilibrium is observed but not the value of
$X$\cite{long}. We also want to stress that this very clear
scenario, with a well defined $T_{\rm eff}$ independent of $t$, is
obtained only if the right procedure with fixed $t$ is used
\cite{Berthier,Berthier2007}. The results are completely different
 if $t_w$ is kept fixed. In such a case $T_{\rm eff}$ is a decreasing
function of $t_w$ \cite{long}. The non-commutability of the  two
procedures, i.e.  either with $t$ or with $t_w$ fixed, shows the
non ergodicity of the phenomenon  and opens a wide discussion on
how to analyze the data in more complex situations, such as those
where $T_{eff}$ is measured in real materials and the procedure
with $t_w$ fixed is often used
\cite{Israeloff,Bellon2001,Ocio,Abou03}.}

In conclusion we have shown that a LC quenched close to the
critical point of the Fr\'eedericksz transition presents aging
features, such as a power law scaling of  correlation functions
and the appearance of a well defined $T_{\rm{eff}}>T$. Most
importantly we show the existence of a well defined $T_{\rm eff}$
if the the right procedure with fixed $t$ is used.  What is very
interesting here is that although we are not exactly at
$\epsilon_0=0$ we observe a large interval of time,$0<t<\tau$
where the predicted aging at critical point can be observed. The
results plotted in fig.\ref{FDTs_t_fixed} agree with those of
ref.~\cite{Berthier,Gambassi1} but are different from those
predicted for mean-field \cite{Gambassi}. This opens the
discussion for further theoretical and experimental developments.
It also shows that the study of the quench at critical point is an
interesting and  not completely understood problem by
itself~\cite{Berthier,Gambassi}.

We acknowledge useful discussion with P. Holdsworth, L. Berthier,
M. Henkel and   M. Pleimling .  This work has been partially
supported by ANR-05-BLAN-0105-01.


\begin{thebibliography}{0}
\bibitem{cell} To check for anchoring and pretilt effects, several
kind of coatings (commercial and home-made) have been used.
 The results are independent on the type of
coating used.
\bibitem{long} More details on this point will be given in a
longer report.
\end{thebibliography}
\end{document}